\documentclass[12pt]{iopart}

\usepackage{iopams}
\usepackage{setstack}

\usepackage{graphicx}
\usepackage{float}
\usepackage{hyperref}

\begin{document}

\title[Unifying PEPS contractions]{Unifying Projected Entangled Pair States contractions}

\author{Michael Lubasch, J Ignacio Cirac and Mari-Carmen Ba\~{n}uls}
\address{Max-Planck-Institut f\"ur Quantenoptik, Hans-Kopfermann-Stra\ss{}e 1, 85748 Garching, Germany.}
\ead{michael.lubasch@mpq.mpg.de}

\begin{abstract}
The approximate contraction of a Projected Entangled Pair States (PEPS) tensor network is a fundamental ingredient of any PEPS algorithm, required for the optimization of the tensors in ground state search or time evolution, as well as for the evaluation of expectation values.
An exact contraction is in general impossible, and the choice of the approximating procedure determines the efficiency and accuracy of the algorithm.
We analyze different previous proposals for this approximation, and show that they can be understood via the form of their environment, i.e.\ the operator that results from contracting part of the network.
This provides physical insight into the limitation of various approaches, and allows us to introduce a new strategy, based on the idea of clusters, that unifies previous methods.
The resulting contraction algorithm interpolates naturally between the cheapest and most imprecise and the most costly and most precise method.
We benchmark the different algorithms with finite PEPS, and show how the cluster strategy can be used for both the tensor optimization and the calculation of expectation values.
Additionally, we discuss its applicability to the parallelization of PEPS and to infinite systems (iPEPS).
\end{abstract}


\maketitle


\newpage

\section{Introduction}

In the last years, Tensor Network States (TNS) have revealed as a very promising choice for the numerical simulation of strongly correlated quantum many-body systems. 
A sustained effort has led to significant conceptual and technical advancement of these methods, e.g.\ \cite{VidalCanForm}-\cite{XiangSRG}.

In the case of one-dimensional systems, Matrix Product States (MPS) are the variational class of TNS underlying the Density Matrix Renormalization Group (DMRG) \cite{RommerDMRG}.
Insight gained from quantum information theory has allowed the understanding of DMRG's enormous success at approximating ground states of spin chains, and the extension of the technique to dynamical problems \cite{VidalCanForm}-\cite{VidalThermal} and lattices of more complex geometry \cite{CiracOriginalPEPS}-\cite{VidalMERA}.

Projected Entangled Pair States (PEPS) \cite{CiracOriginalPEPS} constitute a family of TNS that naturally generalizes MPS to spatial dimensions larger than one and arbitrary lattice geometry.
As MPS, PEPS incorporate the area law by construction, what makes them a very promising variational ansatz for strongly correlated systems which might not be tractable by other means, e.g.\ frustrated or fermionic states where Quantum Monte Carlo methods suffer from the sign problem.
Although originally defined for spin systems, PEPS have been subsequently formulated for fermions \cite{EisertFermions}-\cite{VidalFermions}, and their potential in the numerical simulation of fermionic phases has been demonstrated \cite{VidalFermions}-\cite{CorbozFermions2}.
But in contrast to MPS, even local expectation values cannot be computed exactly in the case of PEPS.
This is because the exact evaluation of the TN that represents the observables has an exponential cost in the system size.
The same difficulty affects the contraction of the TN that surrounds a given tensor, the so-called \emph{environment}, required for the local update operations in the course of optimization algorithms.
It is nevertheless possible to perform an approximate TN contraction with controlled error, albeit involving a much higher computational cost than in the case of MPS.
This limits the feasible PEPS simulations to relatively small tensor dimensions.

Lately, several algorithmic proposals have come out \cite{JiangSU, PizornSL, WangMC, WangCU} that make larger tensors accessible by using new approximations in the PEPS contraction.
Although these approaches allow the manipulation of a larger set of PEPS, their assumptions have an impact on the accuracy of the ground state approximation, and this accuracy is not always directly related to the maximum bond dimension the algorithm can accommodate.
It is nevertheless possible to analyze the various approximations from the unifying point of view of how they treat the environment contraction, which in turn has a physical meaning.
This allows us to understand how a given strategy may attain only a limited precision approximation to the ground state, even when its computational cost allows for large bond dimensions.

Contraction strategies proposed in the literature include the original PEPS method \cite{CiracOriginalPEPS}, the Simple Update \cite{JiangSU}
\footnote{In the original proposal \cite{JiangSU} the Simple Update does not denote a contraction strategy but a tensor update procedure for imaginary time evolution.
However, the environment used in this update corresponds also to a certain contraction method, as we will show later.}
and the Single-Layer \cite{PizornSL} algorithm.
In this work we investigate these algorithms from the unifying perspective introduced above, and present a new contraction scheme that naturally interpolates between the cheapest and most imprecise method and the most expensive and precise one.
We illustrate our findings with finite-size PEPS with open boundary conditions.
A finite PEPS, in which each tensor contributes independent variational parameters, is less biased than its infinite counterpart iPEPS, in which a unit cell of variational tensors is replicated infinitely often and the form of that unit cell can have an effect on the observed order.
However, all our results apply also to iPEPS, and, as we will argue, provide the basis for a new promising approach in that context.

This article is structured as follows.
In section \ref{sec:PEPS} we briefly introduce the basic PEPS concepts and original algorithms.
Section \ref{sec:SU} reviews the Simple Update method introduced in \cite{JiangSU}, and analyzes its performance with finite PEPS.
We find that the resulting ground state energies can be less accurate than those of the original algorithm when the environment form assumed in the method is far from the true one.
The Single-Layer algorithm proposed in \cite{PizornSL} can be seen as a first, conceptual generalization of the Simple Update, and we investigate it in section \ref{sec:SL}.
We show that the error introduced by this method exhibits a strong system size dependence, in contrast to the original algorithm.
With the gained insight, in section \ref{sec:CU} we formulate and investigate a new strategy for the environment approximation based on the idea of clusters, that is applicable to both the tensor update and the computation of expectation values.
Furthermore, we discuss how this cluster strategy is also beneficial to the parallelization of PEPS as well as to the infinite case, i.e.\ iPEPS.
Finally in section \ref{sec:Conclusions} we briefly summarize our results.

\section{PEPS: basic concepts and algorithms}
\label{sec:PEPS}

\begin{figure}[H]
\centering
\includegraphics[width=0.63\textwidth]{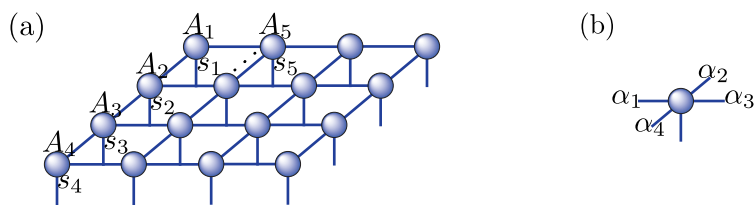}
\caption{\label{fig:PEPS}
                (a) A $4 \times 4$ PEPS $|\psi^{PEPS}\rangle := \sum_{s_{1}, s_{2}, \ldots, s_{16}} \mathcal{F}(A_{1}^{s_{1}} A_{2}^{s_{2}} \ldots A_{16}^{s_{16}}) |s_{1} s_{2} \ldots s_{16}\rangle$ on a square lattice.
                (b) A tensor from the interior with four virtual indices $\alpha_{1}$ to $\alpha_{4}$.
                The function $\mathcal{F}$ performs the contraction of the tensor network by summing over connected virtual indices.
               }
\end{figure}

We consider a system of $N$ quantum particles with Hilbert spaces of dimensions $d_{l}$, for  $l = 1, \ldots, N$, spanned by individual bases $\{|s_{l}\rangle\}$, with $s_{l} = 1, \ldots, d_{l}$.
Projected Entangled Pair States (PEPS) \cite{CiracOriginalPEPS} are states for which the coefficients in the product basis are given by the contraction of a tensor network,
\begin{eqnarray*}
 |\psi^{PEPS}\rangle := \sum_{s_{1}, s_{2}, \ldots, s_{N}} \mathcal{F}(A_{1}^{s_{1}} A_{2}^{s_{2}} \ldots A_{N}^{s_{N}}) |s_{1} s_{2} \ldots s_{N}\rangle \qquad ,
\end{eqnarray*}
with one tensor $A_{l}$ per physical site.
The tensors $A_{l}$ are arranged in a certain lattice geometry and connected to neighboring sites by shared indices, such that the coordination number, $c$, of a certain lattice site coincides with the number of connecting indices.
The latter are called virtual, and apart from them, each tensor $A_{l}$ possesses one physical index $s_{l}$, standing for the physical degree of freedom of the quantum particle on lattice site $l$.
The function $\mathcal{F}$ represents the contraction of all virtual indices.
Each of them ranges up to the parameter $D$ which is named bond dimension.
$D$ determines the number of variational parameters of each tensor, namely $dD^{c}$
\footnote{In the case of open boundary conditions the tensors on the boundaries have fewer virtual indices and variational parameters.}.
The bond dimension sets an upper bound to the entanglement entropy of the state, in fulfillment of the area law.
In particular, if we consider a subsystem delimited by a regular shape of side length $\ell$, the entropy of its reduced density matrix, $\rho_{\ell}$, is upper-bounded by $S(\rho_{\ell})_{\mathrm{max}} \propto \ell^{dim-1} \log(D)$, where $dim$ denotes the system's dimensionality.
Throughout this work, we consider PEPS on two-dimensional square lattices of size $N = L \times L$ with side length $L$ and open boundary conditions.
An example is shown in figure \ref{fig:PEPS}.

\begin{figure}
\centering
\includegraphics[width=0.63\textwidth]{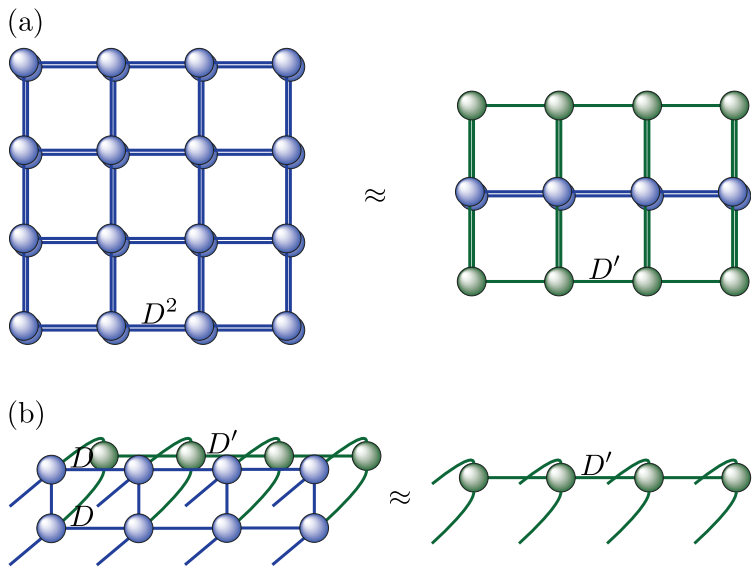}
\caption{\label{fig:Sandwich}
                Sandwich contraction of the norm $\langle \psi | \psi \rangle$.
                (a) The TN corresponding to $\langle \psi | \psi \rangle$ results from figure \ref{fig:PEPS} (a) as ket and as bra and contraction over the physical indices (left).
                      The hard computation of this two-dimensional PEPS sandwich is approximated by an efficiently contractible one-dimensional MPS expectation value (right).
                (b) This is done by successively approximating the action of a bulk row of the sandwich, of bond dimension $D$, on a boundary MPO, of bond dimension $D'$, (left) by a new boundary MPO (right).
                      The latter can be determined with computational cost $\Or(D^{4}D'^{3})+\Or(dD^{6}D'^{2})$.
               }
\end{figure}

For general PEPS, the computation of an expectation value or even the norm is known to be hard \cite{SchuchPEPSComplexity}, like the evaluation of a two-dimensional classical partition function \cite{NishinoCTM}.
Hence only an approximate contraction is possible for already moderate lattice sizes.
The originally proposed algorithm \cite{CiracOriginalPEPS} approximates the two-dimensional TN of an expectation value $\langle \psi | \hat{O} | \psi \rangle$ or the norm $\langle \psi | \psi \rangle$ by means of a succession of one-dimensional MPS contractions, as sketched in figure \ref{fig:Sandwich} (a) for the norm.
In the following we refer to this original method by the term \emph{sandwich contraction}.
The procedure starts by identifying two opposite sides of the TN, e.g.\ the upper- and bottommost rows, with MPS, and each of the intermediate rows with a Matrix Product Operator (MPO) \cite{VerstraeteMPO}.
Beginning from one of the edges, the contraction of the last row with the immediately neighboring one is then a MPS-MPO product, which can be optimally approximated by a MPS of fixed bond dimension, $D'$.
By repeating the procedure from both opposite sides, successive MPS-MPO approximations lead to a representation of both halves of the TN by MPS.
Finally the row in the center is contracted between both MPS to give the approximate expectation value or norm.

At each point of this procedure, the obtained MPS approximates the boundary between the contracted part of the network and the rest.
This MPS can be interpreted as an operator that maps the virtual indices of the ket boundary to the bra and thus we will refer to it as the \emph{boundary MPO}, shown in figure \ref{fig:Sandwich} (b)
\footnote{If the contraction was exact, this boundary MPO would always be positive in the case of the norm computation, due to the bra-ket structure of all the rows involved, but this positive character is in general lost in the truncation.}.
The approximate contraction of the norm has the leading cost $\Or(D^{4}D'^{3})+\Or(dD^{6}D'^{2})$, and thus both cost and error are determined solely by the bond dimension $D'$ of the boundary MPO.
Although in principle $D'$ could scale exponentially with the number of rows, in practice it typically scales as $D' \propto D^{2}$ independent of the system size, such that the original sandwich contraction has the total computational cost $\Or(D^{10})$.

For certain problems, this observed mild scaling can be given a more rigorous ground.
Indeed, the boundary MPO can be interpreted as the thermal state of a Hamiltonian defined on the virtual degrees of freedom of the boundary.
This boundary Hamiltonian is obtained by identifying its excitation spectrum with the entanglement spectrum of the state \cite{LiHaldane}.
Such a construction is very natural in the framework of PEPS and establishes a holographic principle \cite{CiracBoundaryTheories}.
While PEPS are expected to represent the low energy sector of local Hamiltonians well, it has not been proven when expectation values can be computed efficiently with them.
However, if the boundary Hamiltonian is local, as evidence suggests for gapped models \cite{CiracBoundaryTheories}, the corresponding thermal state will be efficiently approximated by a MPO \cite{Hastings}.

In the following, we obtain the PEPS approximation to the ground state of a certain Hamiltonian by means of imaginary time evolution.
It is based on the idea that $e^{- t \hat{H}} |\psi_{0}\rangle$ converges to the ground state of $H$ exponentially fast with $t$, as long as the ground state is not degenerate and has non-vanishing overlap with the initial state, $|\psi_{0}\rangle$.
In the context of TNS \cite{VidalTE}, the initial state is chosen within the appropriate TNS family, and a Suzuki-Trotter decomposition of the evolution operator $U(t) = e^{- t \hat{H}} = ( e^{- \tau \hat{H}} )^n$ is applied to local Hamiltonians, such that each step of the evolution, $\tau=t/n$, is approximated by a product of local \emph{Trotter gates}.
The resulting state after each gate or set of gates, is again approximated by an adequate TNS.
In particular, the action of a certain operator $\hat{O}$ on a PEPS $|\phi\rangle$ can be approximated by a new PEPS $|\psi\rangle$ by minimizing the cost function $\mathrm{d}(|\psi\rangle) = |||\psi\rangle - \hat{O}|\phi\rangle||^{2}$.
We perform this minimization for each gate via an alternating least squares (ALS) scheme, optimizing one tensor at a time while the others are fixed, and sweeping only over the tensors on which the Trotter gate acts.
The optimal tensor at position $l$ is the solution of a system of linear equations $N_{l} \bi{A_{l}} = \bi{b_{l}}$, where the norm matrix $N_{l}$ is defined from the tensor network $\langle \psi | \psi \rangle$ by leaving out the tensor $A_{l}$ in the ket and $A_{l}^{*}$ in the bra, and the vector $\bi{b_{l}}$ results from the tensor network $\langle \psi | \hat{O} | \phi \rangle$ by removing $A_{l}^{*}$ from the bra.

The environment of a tensor at site $l$ is the open TN that results when this tensor and its adjoint are removed from the norm of the state.
Contracting the environment is necessary to evaluate $N_{l}$ and $b_{l}$, which determine the local equation for $A_{l}$.
Such contraction can only be carried out approximately, and the approximation strategy is decisive both for the accuracy and for the computational cost of the algorithm
\footnote{In general, also the tensor update operations contribute to the final cost.
                 If we restrict the variational parameters to the reduced tensor \cite{VidalFermions, WangMC}, the linear equations can be solved with $\Or(D^{6})$ operations.}.
Because we process the Trotter gates one after another and modify only the tensors on which the gate directly acts, in the following we focus on the contraction of the norm TN around a Trotter gate.
The importance of the environment approximation has been recognized also in other works, e.g.\ \cite{Pfeifer}, or in the different context of tensor renormalization group algorithms \cite{NaveTRG} where a more precise environment representation lead to significant improvements \cite{XiangSRG}.

\section{Simple Update}
\label{sec:SU}

\begin{figure}[H]
\centering
\includegraphics[width=0.815\textwidth]{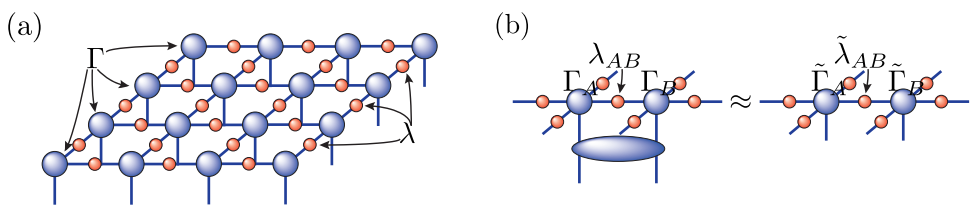}
\caption{\label{fig:SU}
                (a) A $4 \times 4$ PEPS $|\psi^{PEPS,SU}\rangle := \sum_{s_{1}, s_{2}, \ldots, s_{16}} \mathcal{F}(\Gamma_{1}^{s_{1}} \lambda_{1} \ldots \Gamma_{16}^{s_{16}}) |s_{1} s_{2} \ldots s_{16}\rangle$ of the Simple Update (SU) form,
                      composed of $\Gamma$ tensors and $\lambda$ matrices.
                (b) Assuming nearest-neighbor Trotter gates, the 6 $\lambda$ matrices surrounding a tensor pair $\Gamma_{A}$ and $\Gamma_{B}$ are sufficient for the update of this pair and its $\lambda_{AB}$.
               }
\end{figure}

The Simple Update method (SU) \cite{JiangSU} directly generalizes the one-dimensional TEBD \cite{VidalCanForm, DaleyTE, VidalTE, WhiteTE, iTEBD} and proposes for the PEPS tensors the decomposition
\begin{eqnarray*}
 |\psi^{PEPS,SU}\rangle & := & \sum_{s_{1}, s_{2}, \ldots, s_{N}} \mathcal{F}(\Gamma_{1}^{s_{1}} \lambda_{1} \Gamma_{2}^{s_{2}} \lambda_{2} \ldots \Gamma_{N}^{s_{N}}) |s_{1} s_{2} \ldots s_{N}\rangle \qquad ,
\end{eqnarray*}
formally analogous to the canonical form for MPS \cite{VidalCanForm}, where the $\Gamma_{l}$ are tensors with the same dimensions as the original $A_{l}$, and the $\lambda_{l}$ are $D \times D$ diagonal and positive matrices, see figure \ref{fig:SU}.
Although in the case of PEPS, the $\lambda$ matrices do not have the clear meaning of their one-dimensional counterparts, the SU has proven a successful strategy in the context of iPEPS, starting with \cite{JiangSU}.
This success can be attributed, on the one hand, to the low computational cost of the tensor update, which is why large values of $D$ can be reached easily.
Indeed, the SU rule requires only the $\lambda$ matrices that are closest to a tensor pair and as a consequence has the computational cost $\Or(D^{5})$.
On the other hand, all parts of the algorithm are well-conditioned.
These positive aspects arise at the expense of an oversimplified representation of the environment as separable and local, that, in general, can only be a rough approximation of the true environment.

In order to illustrate its performance, we employ the SU to find ground states of the Quantum Ising Hamiltonian with transverse field
\begin{equation}
 \hat{H} =  - \sum_{\langle l, m \rangle} \sigma_{l}^{z} \otimes \sigma_{m}^{z} - B \sum_{l} \sigma_{l}^{x}
\label{eq:IsingH}
\end{equation}
and of the Heisenberg model
\begin{equation}
 \hat{H}  =  \sum_{\langle l, m \rangle} \vec{S}_{l} \cdot \vec{S}_{m} \qquad ,
\label{eq:HeisenbergH}
\end{equation}
where $\langle l , m \rangle$ denotes pairs of neighboring sites $l$ and $m$.
In the context of finite PEPS considered here, to the best of our knowledge, the SU had not been yet used.
We determine the ground state of a particular problem by evolving an initial state long enough in imaginary time and successively decreasing the time step $\tau$ until convergence.

\begin{figure}[H]
\centering
\includegraphics[width=0.63\textwidth]{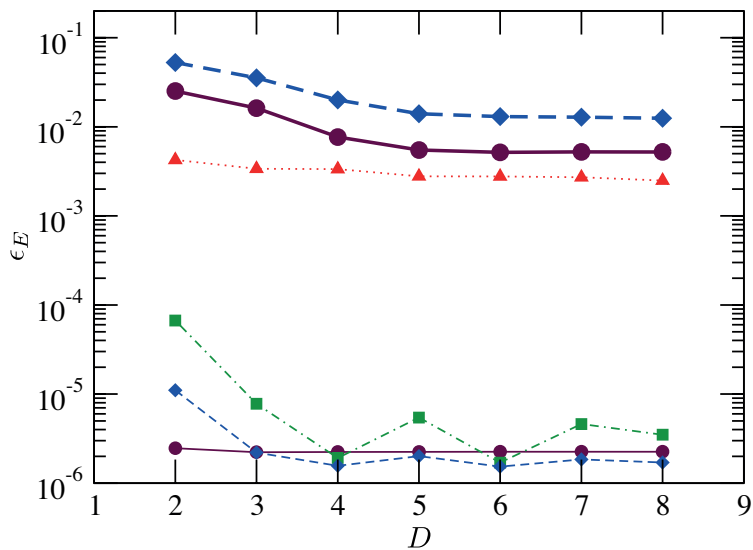}
\caption{\label{fig:DESU}
                Relative energy error $\epsilon_{E} := |E(D)-E_{0}|/|E_{0}|$ of the SU, $E(D)$, with regard to the exact ground state energy, $E_{0}$.
                We consider the Ising model (thin lines) on a $4 \times 4$ lattice with $B=1.0$ (solid), $2.0$ (dotted), $3.0$ (dash-dotted), $4.0$ (dashed), and the Heisenberg model (thick lines) on a $4 \times 4$ (dashed) and $10 \times 10$ (solid) lattice.
               }
\end{figure}

Figure \ref{fig:DESU} compares SU results to exact ground state energies.
The scheme performs remarkably well for the Ising model at $B=1.0$, $3.0$, and $4.0$, where the relative energy error is below $10^{-5}$ already with $D=3$.
But at $B=2.0$ and for the Heisenberg Hamiltonian, we observe that the energy does not improve significantly beyond a certain value of the bond dimension, and remains far from the exact value.
We identify this as a limitation, not of the ansatz, but of the update procedure, since the original PEPS algorithm \cite{CiracOriginalPEPS} achieves for the Heisenberg model on a $4 \times 4$ lattice with $D=3$ already lower energy than any of the SU values from the figure, and with $D=4$ it attains an energy per site $-0.5739$ already very close to the exact value $-0.5743$.
Although we observed that the SU result can depend on the initial state, in particular for the larger bond dimensions
\footnote{This became evident by running the algorithm with various values of $D$ separately, each run starting from a product state in which the tensors' zeroes were replaced by random numbers from the interval $[-0.01, 0.01]$.
                 In this setting we observed that the SU can lead to a final $D=8$ energy above the $D=7$ value.},
this dependence appeared not so strongly when we increased the bond dimension successively during imaginary time evolution.

\section{Single-Layer}
\label{sec:SL}

The Single-Layer (SL) algorithm for the computation of the norm $\langle \psi | \psi \rangle$ was presented in \cite{PizornSL}.
This method takes into account the bra-ket structure of the sandwich, and maintains it and hence the positive character of the environment while the contraction of the network progresses from one edge.
To achieve this structure, the approximate contraction proceeds by successive MPO-MPS operations, like the original algorithm, but this time performed on a single layer of the sandwich TN.
Then the boundary, i.e.\ the already contracted part of the network, is always approximated by a purification MPO \cite{VerstraeteMPDO}, namely the result of tracing out a part of the physical indices at every site of a MPS.
This MPS is assumed to have some maximum bond dimension, $D''$, and physical dimension $D \times d'$, where $d'$ is the dimension of the traced out degrees of freedom, what we call \emph{purification bond}.
In this way, the local and separable environment defined by the $\lambda$ matrices in the SU is generalized by means of purification MPS that can better capture non-local and non-separable boundary correlations.
Moreover, the boundary purification MPO is always a positive operator, and it allows to devise a stable tensor update procedure for imaginary time evolution \cite{PizornSL}.

\begin{figure}[H]
\centering
\includegraphics[width=0.815\textwidth]{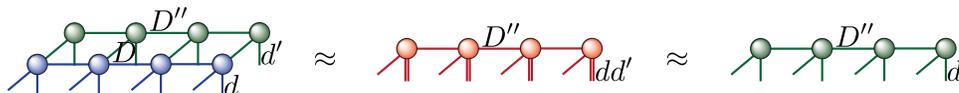}
\caption{\label{fig:SL}
                SL contraction of the norm $\langle \psi | \psi \rangle$.
                As explained in the text, the approximation of figure \ref{fig:Sandwich} is achieved by means of operations in the ket alone.
               }
\end{figure}

The SL operations take place in the two steps shown in figure \ref{fig:SL}.
First, the ket part of a PEPS row is applied as a MPO to the MPS of the boundary purification.
The result is truncated to a MPS with bond dimension $D''$ and increased purification bond, $dd'$.
Then the purification bond is reduced from $dd'$ to $d'$, by imposing the canonical form \cite{VidalCanForm} and projecting the reduced density matrix of each site onto the space spanned by its $d'$ largest eigenvectors.
The computational cost of the first step, which proceeds via the standard ALS scheme, scales as $\Or(dd'D^{4}D''^{2})+\Or(dd'D^{2}D''^{3})$, while the leading cost of the second step is $\Or(d^{2}d'^{2}DD''^{2})$, negligible only when $d'$ is small.  
Because the purification bond satisfies $d' \leq DD''^{2}$, the maximum cost can at most grow as $\Or(dD^{5}D''^{4})+\Or(dD^{3}D''^{6})$ when $d'$ takes on its largest possible value.
In \cite{PizornSL, WangCU} it was proposed to set $d'=D=D''$, in which case the number of operations scales only like $\Or(D^{7})$, and a clear computational gain compared to the original contraction can be expected.

In order to analyze the performance of the SL procedure, we study the accuracy of the norm contraction, $\mathcal{N} \equiv \langle \psi | \psi \rangle$, as a function of the truncation parameters, $D''$ and $d'$, for a set of different PEPS, and compare the results to those of the original algorithm.
In particular, we consider $D=2-4$ PEPS ground state approximations from the SU for the Ising model (\ref{eq:IsingH}) with transverse fields $B=1.0$, $2.0$, $3.0$, and $4.0$, on lattices with side lengths $L=11$ and $21$.
In all cases, the exact norm was estimated by means of the sandwich contraction with bond dimension $D'=100$, large enough to make the error negligible.

In the case of the original algorithm, the relative error always decreases exponentially with the bond dimension of the general boundary MPO, $D'$.
Moreover, for a fixed bond dimension of the PEPS, $D$, this error shows no system size dependence.
In the SL algorithm, for fixed purification bond $d'$, the contraction error converges quickly as function of $D''$ to a final value that is entirely determined by $d'$.
Even when that purification bond dimension takes on its maximum value, $d'=DD''^{2}$, this error lies many orders of magnitude above the one from the sandwich contraction with the same $D'=D''$.
It is worth noticing that for large $D' = D'' \gg D$ the computational cost of the original method is actually lower than the one of the SL algorithm with maximum $d'=DD''^{2}$.

\begin{figure}[H]
\centering
\includegraphics[width=0.63\textwidth]{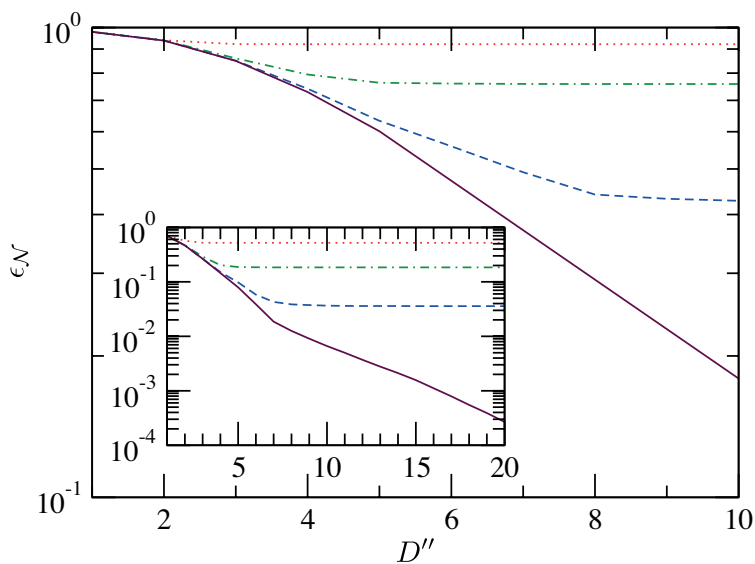}
\caption{\label{fig:DNSL}
                Relative norm error $\epsilon_{\mathcal{N}}$ in the SL contraction of a $D=2$ SU ground state approximation of the Ising model with $B=3.0$ on a $11 \times 11$ (inset) and $21 \times 21$ (main) lattice.
                We consider $d'=2$ (dotted), $d'=4$ (dash-dotted), $d'=8$ (dashed), and the maximum possible $d'=DD''^{2}$ (solid).
                We defined $\epsilon_{\mathcal{N}} := |\mathcal{N}(D'')-\mathcal{N}_{ref}|/|\mathcal{N}_{ref}|$ where the reference value $\mathcal{N}_{ref}$ was computed with the sandwich contraction using $D'=100$.
               }
\end{figure}

The differences between the original and the SL contraction become even more apparent when the lattice size is increased to $N = 21 \times 21$, because the SL algorithm depends strongly on the system size as can be gathered from Fig.~\ref{fig:DNSL}.
In that case, given $d'=DD''^{2}$ and $D''=10$, the norm error grows from $\epsilon_{\mathcal{N}} \approx 0.007$ in the $11 \times 11$ to $\epsilon_{\mathcal{N}} \approx 0.1$ in the $21 \times 21$ lattice, in marked contrast to $\epsilon_{\mathcal{N}} \approx 10^{-11}$ in the sandwich contraction with $D'=10$ obtained for both lattice sizes.
And we observe a similar scaling for PEPS with larger bond dimensions.
For instance, the SL value to $D=4$, $d'=8$, and $D''=10$ grows from $\epsilon_{\mathcal{N}} \approx 0.06$ in the $11 \times 11$ to $\epsilon_{\mathcal{N}} \approx 0.6$ in the $21 \times 21$ lattice, which has to be compared to $\epsilon_{\mathcal{N}} \approx 10^{-5}$ in the sandwich contraction with $D'=10$ achieved for both lattice sizes.

From this analysis we conclude that the choice $d' = D = D''$, which ensures the advantageous computational cost $\Or(D^{7})$, is in general too restrictive in order to get a comparable precision to that of the original algorithm.
Moreover, because the required values of the parameters $d'$ and $D''$ for a certain contraction precision depend strongly on the system size, one cannot make a general statement about the cost scaling of the SL algorithm.
This is different from the situation in the original algorithm, where the parameter $D'$ controlling the cost can typically be chosen as $D' \propto D^{2}$ with a prefactor that seems not to depend on the system size but only on the state.

The environment approximation in the SL scheme, despite being positive, does not correspond to the most general boundary purification, a fact that provides some insight into the limitations of the method.
The purification of a mixed state is only defined up to an isometry on the traced-out degrees of freedom.
But the optimization in the SL algorithm does merely allow for local isometries, i.e.\ for tensor products of isometries each acting on a single site of the boundary only.
It is possible to devise an ALS algorithm that searches the optimal general boundary purification, at the expense of a cost function for each site which is no longer quadratic but quartic in the local tensor variables.
The minimum of such a cost function is no longer the solution of linear but of nonlinear equations, which are numerically much more demanding than the simple QR decomposition that gives the optimal general boundary MPO in the original algorithm.
Therefore such a strategy may result in an undesirable slowing down of the algorithm.
Notice also that, while a given purification can always be written efficiently as a positive MPO, namely via contraction of the tensors at each site over their purification bond, the reverse statement is not true, since there exist positive MPO that cannot be written efficiently as purifications \cite{Gemma}.
We conclude that, for the problems considered here, it is more advisable to work with a general boundary MPO upon which positivity is not explicitly imposed
\footnote{Although, when $D'$ is chosen too small, the negative eigenvalues of the environment can lead to instabilities in the tensor update, when $D'$ is large enough, the tensor update is stable and then more accurate.},
and based on it formulate contraction algorithms where cost and precision grow simultaneously, as we shall do in the following section.

\section{Clusters}
\label{sec:CU}

The most precise environment approximation is achieved by the original algorithm, in the form of a MPO with sufficiently large bond dimension $D'$.
On the opposite extreme of the spectrum, the lowest computational cost corresponds to the SU, where the environment is represented by a tensor product of matrices each acting on a single virtual bond only.
Here, we aim at a contraction scheme that allows to systematically tune the environment precision together with the cost and that interpolates between the SU and the original algorithm.

This goal is achieved with the help of \emph{clusters}.
In a state with short-range correlations, we expect that the major contribution to the environment of a given tensor comes from the closest sites.
If such sites are not correlated with further ones, or among themselves, the environment will actually be a product, similar to the SU approximation. 
Correlations in the state cause the environment to be non-separable in general, and to incorporate relevant contributions from faraway sites.
Hence, by progressively taking into account the contribution of sites at longer distances, we would improve the environment description.

In our PEPS algorithm, we are interested in the environment of a row (respectively column), which is required for the update of all the tensors in it.
We therefore define a \emph{cluster} of size $\delta$ around a certain row as all the surrounding rows which are separated from it by a distance smaller or equal to $\delta$, and similarly for columns.
The environment contribution from sites outside the cluster can be roughly approximated by a product in the spirit of the SU, while the contribution from sites inside the cluster is taken into account with more precision, as in the original algorithm.
This defines a new contraction scheme that we call Cluster Update in analogy to \cite{WangCU}, and that we abbreviate as CU$_{\delta}$ for cluster size $\delta$.
The limiting cases of this strategy are $\delta = L-1$, when the environment reverts to the one of the original algorithm, and $\delta = 0$, which is closely related to the SU.

\subsection{Cluster size $\delta=0$: a generalized Simple Update}

\begin{figure}[H]
\centering
\includegraphics[width=0.63\textwidth]{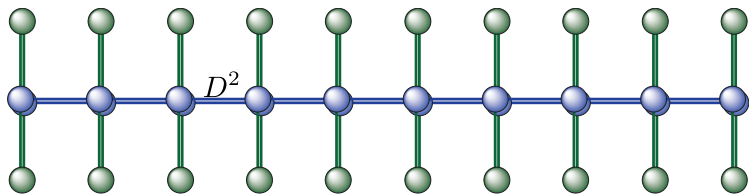}
\caption{\label{fig:SepEnv1}Separable Environment with $D'=1$.}
\end{figure}

\begin{figure}[H]
\centering
\includegraphics[width=0.63\textwidth]{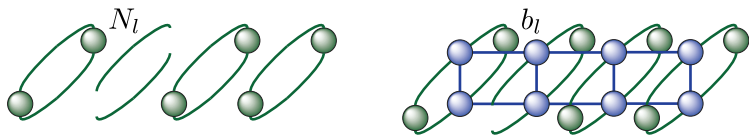}
\caption{\label{fig:SepEnv2}
                Tensor contractions during the ALS sweeping in the computation of the environment for CU$_{0}$.
                For each local update, the norm matrix (left) is the identity times a positive constant, such that the solution is simply proportional to $b_{l}$ (right), which is given by a positive TN if each of the local tensors is positive.
               }
\end{figure}

The particular case $\delta = 0$ (CU$_{0}$) leads to the environment approximation of a certain row, or column, as a product MPO, illustrated in figure \ref{fig:SepEnv1}.
This can be found by optimizing the boundary MPO with $D'=1$, where the standard MPO-MPS ALS scheme can now yield a positive MPO.
Indeed, if each of the local tensors of the MPO is positive, this positivity is maintained during the update procedure, since for each local optimization the norm matrix in $N_{l} \bi{A_{l}} = \bi{b_{l}}$ is proportional to the identity, and the TN to $b_{l}$ is positive, as explained in figure \ref{fig:SepEnv2}.
Starting the ALS sweeping from an initial positive MPO, which can be trivially constructed from positive local tensors (e.g.\ of the form $X^{\dag}X$ with random $X$), ensures then a positive environment.
Moreover, all contractions can be performed with $\Or(dD^{5})$ operations, so that the computation of the optimal separable environment does not exceed the leading cost of the SU.

Imaginary time evolution based on a positive separable boundary MPO leads to an algorithm which is very similar to the SU.
Both schemes are characterized by the same computational cost
\footnote{Just like the environment approximation, the tensor update in a separable environment can be performed with $\Or(D^{5})$ operations.}
and make use of a separable environment, but the CU$_{0}$ method proposed here optimizes the approximate boundary over all possible separable MPO, and hence can be interpreted as a generalized Simple Update.

In order to elucidate the connection between both algorithms, we study how imaginary time evolution with CU$_{0}$ changes PEPS ground state approximations from the SU for the Ising model (\ref{eq:IsingH}) with various magnetic fields.
In our quantitative comparison we consider a specific virtual bond between two neighboring sites in the center of the lattice, and focus on the corresponding $\lambda$ matrix generated by the SU after convergence, $\lambda^{\mathrm{SU}}$.
The diagonal of that matrix can directly be compared to the converged singular values emerging in the CU$_{0}$ every time a gate is applied to this particular pair of sites, $\lambda^{\mathrm{CU}_{0}}$.
As shown in table \ref{tab:GSU} for a $11 \times 11$ PEPS with $D=2$, the relative difference between the entries of these two $\lambda$ matrices is below $\approx 10^{-2}$.

We can analyze the similarities between both algorithms in more detail by looking at the role of the $\lambda$ matrices in the environment for the update operations.
In the SU, the entries in $\lambda^{\mathrm{SU}}$ are determined after applying one gate to the relevant pair of tensors, but (in the here considered case of nearest neighbor interactions) they are not affected by gates which involve only one member of the pair. 
For the latter tensor updates, the $\lambda^{\mathrm{SU}}$ matrix enters the environment unchanged, even after the $\Gamma$ tensors of the pair have been modified.
In contrast, in the CU, the environment for a given update operation depends on the surrounding tensors, and changes every time they are updated.
In the case of CU$_{0}$, a similar role to that of $\lambda^{\mathrm{SU}}$ is played by the eigenvalues of the local tensor in the boundary MPO at the site corresponding to this particular virtual bond, $\Sigma^{\mathrm{CU}_{0}}$.
For nearest neighbor interactions and the bond we are considering, there are six Trotter gates in each time step (see the figure in table \ref{tab:GSU}) that involve only one of the tensors of the pair.
The $\Sigma^{\mathrm{CU}_{0}}$ entries change only slightly, $\approx 10^{-2}$, for each such tensor update, as can be appreciated in table \ref{tab:GSU}.
And their difference to the corresponding $\lambda^{\mathrm{SU}}$ is of the same order.
Additionally, we computed the separable boundary MPO for the SU PEPS and compared the eigenvalues of the local tensors to the corresponding $\Sigma^{\mathrm{CU_{0}}}$, to find a similar agreement.
We observed the same behavior in larger lattices, with larger bond dimensions, as well as on different virtual bonds.

\begin{table}[H]
 \centering
 \caption{\label{tab:GSU}
                 We apply the SU with $D=2$ to a $11 \times 11$ Ising model with different magnetic fields $B$.
                 All $\lambda$ matrices have converged to machine precision and we report the final second entry $\lambda_{2}^{\mathrm{SU}}$ on the vertical virtual bond at row $5$ and column $6$.
                 The resulting PEPS is further evolved with the CU$_{0}$ until convergence.
                 We show the second singular value $\lambda_{2}^{\mathrm{CU_{0}}}$ emerging in the tensor update on the considered bond and the second eigenvalue $\Sigma_{2}^{\mathrm{CU_{0}}}$ of the boundary MPO matrix at that place whenever it enters a tensor update.
                 This happens on the six different positions relative to a tensor pair defined in the figure on the right, during the approximation of the four sets of Trotter gates in one time evolution step.
                 We adopt the normalization in which each first $\lambda$ entry, singular value and eigenvalue is always $1$.
                }
 \begin{indented}
 \item[]\begin{tabular}{@{}lllll}
 \br
                                                                                      & $B=1.0$ & $B=2.0$ & $B=3.0$ & $B=4.0$ \\
 \mr
  $\lambda_{2}^{\mathrm{SU}}$                              & 0.006007 & 0.026032 & 0.078572 & 0.071486 \\
  $\lambda_{2}^{\mathrm{CU_{0}}}$                      & 0.006022 & 0.026252 & 0.078953 & 0.071210 \\
  $\Sigma_{2}^{\mathrm{CU_{0}}}(\mathrm{i})$   & 0.006008 & 0.026049 & 0.078399 & 0.071216 \\
  $\Sigma_{2}^{\mathrm{CU_{0}}}(\mathrm{ii})$  & 0.006008 & 0.026048 & 0.078421 & 0.071216 \\
  $\Sigma_{2}^{\mathrm{CU_{0}}}(\mathrm{iii})$ & 0.005784 & 0.025236 & 0.077678 & 0.071391 \\
  $\Sigma_{2}^{\mathrm{CU_{0}}}(\mathrm{iv})$ & 0.005784 & 0.025237 & 0.077828 & 0.071391 \\
  $\Sigma_{2}^{\mathrm{CU_{0}}}(\mathrm{v})$  & 0.005567 & 0.024434 & 0.076890 & 0.071566 \\
  $\Sigma_{2}^{\mathrm{CU_{0}}}(\mathrm{vi})$ & 0.005567 & 0.024434 & 0.077094 & 0.071566 \\
 \br
 \end{tabular}
 \hspace{6mm}
 \includegraphics[width=0.175\textwidth]{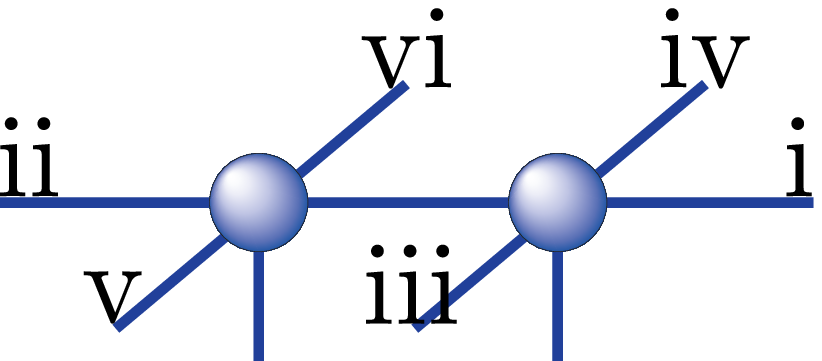}
 \end{indented}
\end{table}

Our observations provide an explanation for the functioning of the SU: because the latter scheme applies the same $\lambda^{\mathrm{SU}}$ matrix to the tensor updates of all four sets of Trotter gates, this $\lambda^{\mathrm{SU}}$ can be seen as a mean value for the six $\Sigma^{\mathrm{CU_{0}}}$ from the optimal positive separable environment, and the SU indeed converges it to that mean value.

The CU with $\delta = 0$ always uses the best separable environment in each tensor update, and therefore depends less on the initial state and can produce energies slightly below the ones from the SU.
However, the final energies of both methods are very close to each other (compare figures \ref{fig:DESU} and \ref{fig:DEL4L10}).

Although the SU is an algorithm completely formulated in the SL, our study in the double layer picture provided crucial insight into it, thus reinforcing the idea that the boundary should be described as a general MPO.

\subsection{From Simple to Full Update}

By considering clusters of size $\delta \geq 1$ we can systematically take into account more of the correlations in the environment approximation.
Outside the cluster, the environment is represented by a separable MPO and determined as in the CU$_{0}$ described above.
Then the cluster tensors are contracted row by row with this boundary, as in the original algorithm, to produce a new boundary MPO with larger bond dimension.
The approximation is controlled by the cluster size and the bond dimension $D'$ used in the contractions within the cluster.

\begin{figure}[H]
\centering
\includegraphics[width=0.63\textwidth]{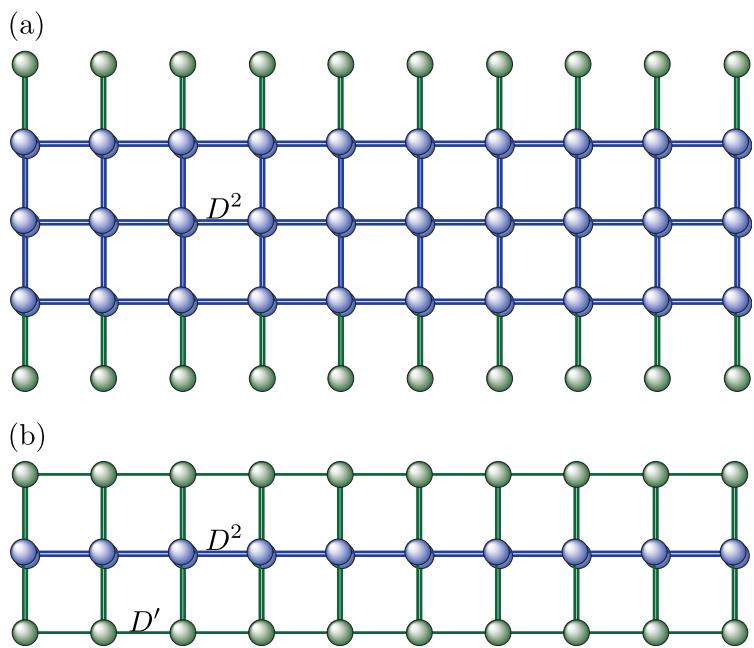}
\caption{\label{fig:C1}
                Environment for the CU$_{1}$ of a middle row.
                (a) Outside the cluster, the approximate contraction of the sandwich is performed via a positive separable boundary MPO.
                (b) The contraction continues inside the cluster via a general boundary MPO with bond dimension $D'$.
               }
\end{figure}

Figure \ref{fig:C1} shows the smallest non trivial cluster, for CU$_{1}$, in which only the two rows adjacent to the one to be updated enter the cluster contraction.
In this case, the optimal boundary MPO with bond dimension $D'$ for the update of a row is computed from the action of a bulk row on a separable boundary MPO with $\Or(dD^{5}D'^{2})$ operations.
This is the dominant cost in the environment approximation of CU$_{1}$, given the fact that the separable MPO outside the cluster is obtained with only $\Or(dD^{5})$ operations.
Hence, the environment approximation for clusters of size $\delta=1$ is computationally cheaper than the full contraction with cost $\Or(D^{4}D'^{3})+\Or(dD^{6}D'^{2})$.

\begin{figure}[H]
\centering
\includegraphics[width=0.63\textwidth]{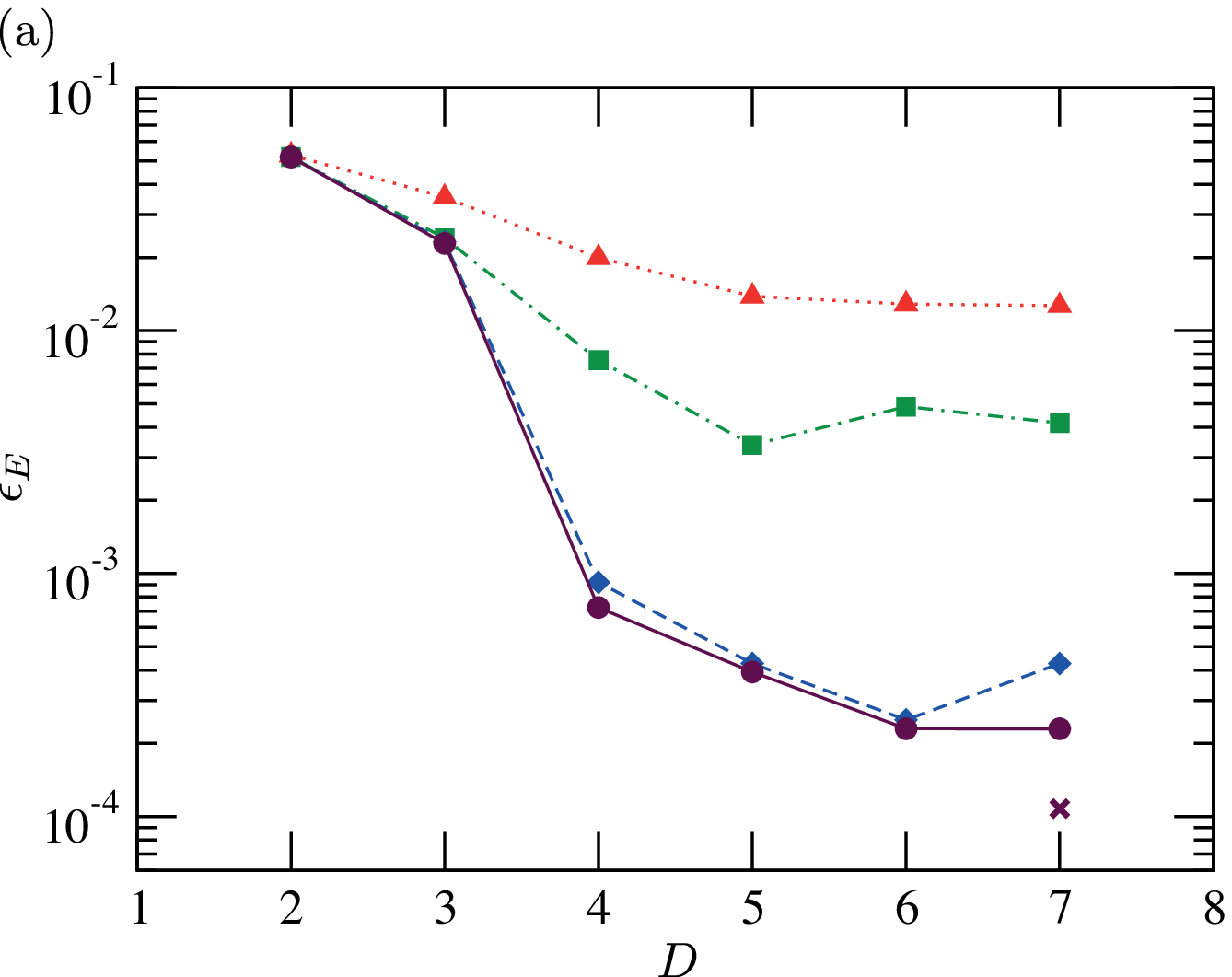}
\includegraphics[width=0.63\textwidth]{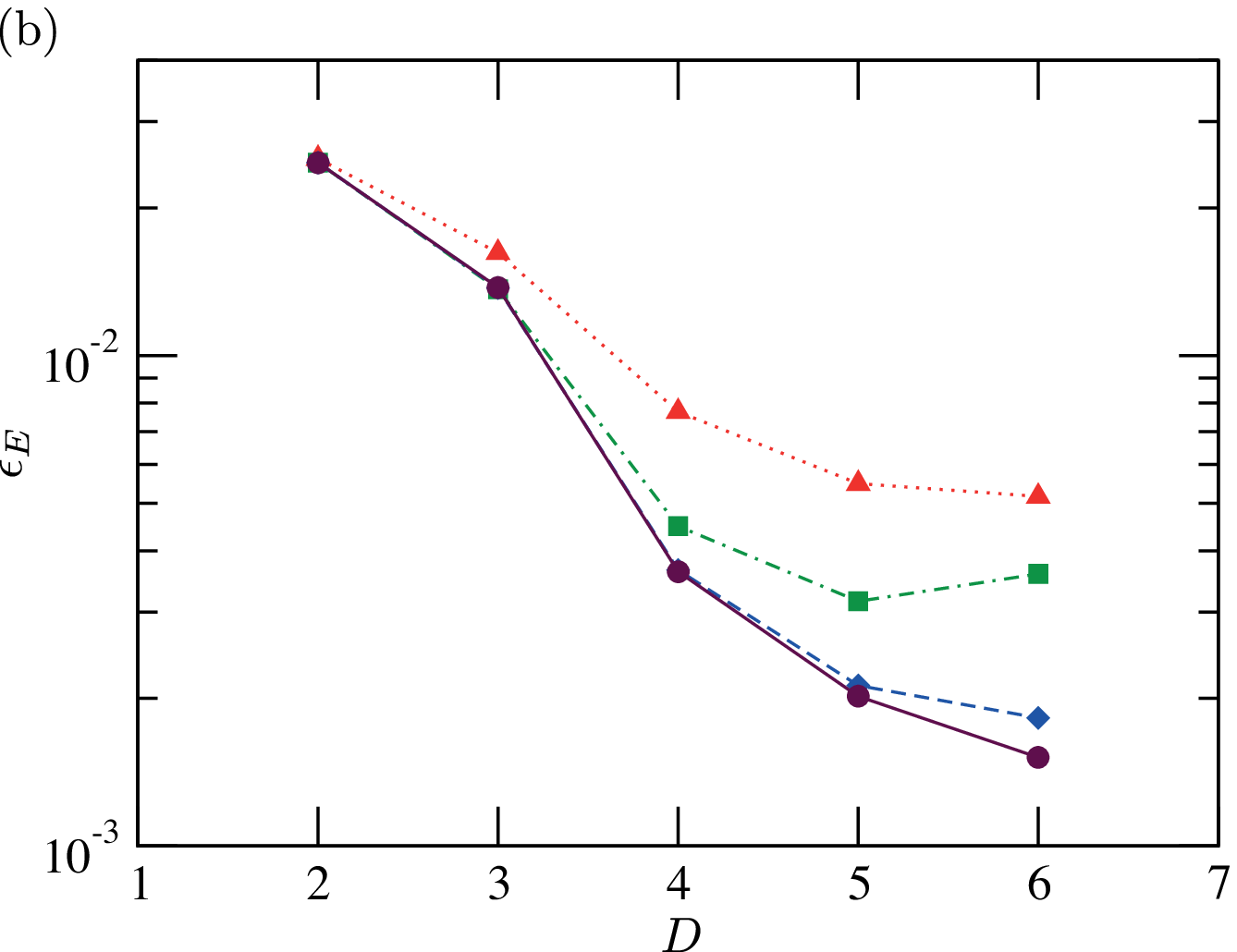}
\caption{\label{fig:DEL4L10}
                Relative energy error $\epsilon_{E}$ as in figure \ref{fig:DESU} for a $4 \times 4$ (a) and $10 \times 10$ (b) Heisenberg model.
                We consider the CU$_{0}$ (dotted), the CU$_{1}$ with $D'=D^{2}$ (dash-dotted), and the FU with $D'=D^{2}$ (dashed), $D'=2D^{2}$ (solid), and $D'=130$ (cross).
               }
\end{figure}

\begin{figure}[H]
\centering
\includegraphics[width=0.63\textwidth]{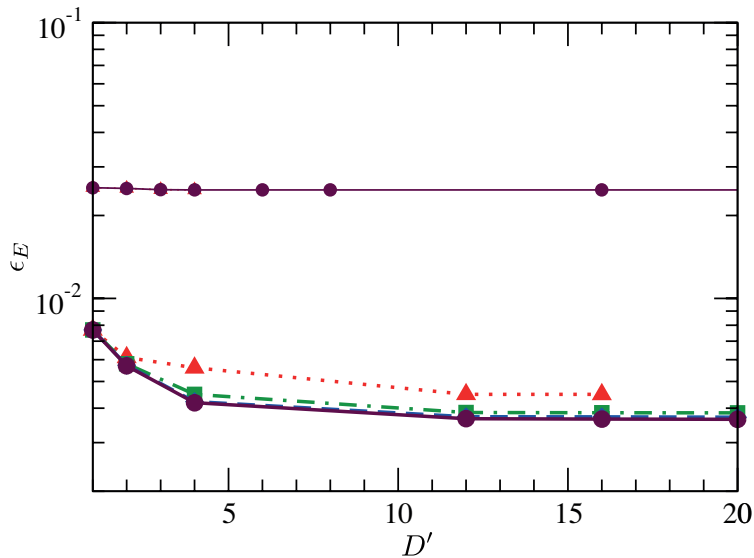}
\caption{\label{fig:DECUL10}
                Relative energy error $\epsilon_{E}$ as in figure \ref{fig:DESU} for a $10 \times 10$ Heisenberg model, obtained with various fixed values of the bond dimension $D'$ of the boundary MPO.
                We propagated $D=2$ (upper thin lines) and $D=4$ (lower thick lines) PEPS with the CU with cluster size $\delta=1$ (dotted), $2$ (dash-dotted), $3$ (dashed), and with the FU (solid).
               }
\end{figure}

To examine the usefulness of this cluster strategy, we compare its performance to that of the SU via their ground state accuracies for the Heisenberg model (\ref{eq:HeisenbergH}).
Starting from converged SU PEPS, we ran the CU imaginary time evolution with several cluster sizes for various bond dimensions $D$ and $D'$ on $4 \times 4$ and $10 \times 10$ lattices.
Figure \ref{fig:DEL4L10} contains our cluster results for $\delta = 0$, $1$, and $L-1$, as function of $D$, such that they can be compared directly to the SU results of figure \ref{fig:DESU}.
The convergence of the CU with cluster size $\delta$ as well as with bond dimension $D'$ can be gathered from figure \ref{fig:DECUL10} for $10 \times 10$ PEPS with $D=2$ and $4$.
We refer to the CU with maximum cluster size $\delta = L-1$ as \emph{Full Update} (FU), a notion taken from iPEPS (see e.g.\ \cite{VidalFermions}).
The FU is not identical to the original PEPS algorithm \cite{CiracOriginalPEPS}, because in the CU the action of single Trotter gates is approximated sequentially, such that for each gate the only tensors updated are those on which the gate directly acts.
Thanks to this procedure, the FU requires just the approximate contraction of the norm sandwich, and is therefore computationally less demanding than the original algorithm, in which, additionally, the PEPS sandwich with a full set of Trotter gates acting on all the state has to be contracted.

We find that the CU$_{0}$ produces very similar energies as the SU, with slight improvement for small systems or for large bond dimensions.
The difference between both methods is most apparent in case of the smaller $4 \times 4$ lattice where the CU$_{0}$ gives lower energies for bond dimensions $D \geq 4$.
This can be understood taking into account that the effect of the system boundary, better captured by the environment approximation in CU$_{0}$, is more important for smaller systems.
We observe then that the CU$_{1}$ improves the SU energies considerably.
Finally, the FU reduces the energy further significantly when $D \geq 4$, and its effect appears more pronounced with increasing bond dimension $D$.
For $D=2$ and $3$, the FU improves upon the CU$_{1}$ only in case of the smaller $4 \times 4$ system.
Notice that the tables \ref{tab:ESU}-\ref{tab:ECUL10D4} contain the precise energy values that were used in this analysis.

From the arguments above it is apparent that a better representation of the environment is crucial for an improved PEPS approximation of the true ground state.
We also infer that larger bond dimensions $D$ require more precise environment representations in the tensor update.
Within the CU, this improvement can be achieved systematically by gradually increasing, firstly, the cluster size $\delta$ and, secondly, for each fixed $\delta$, the boundary bond dimension $D'$.
Indeed, we can see in figure \ref{fig:DECUL10} for each fixed cluster size that with growing $D'$ the energy decreases consistently, as the precision of the environment representation in the tensor update increases.
The energy converges at a certain value $D_{max}'$ that depends both on the bond dimension $D$ of the considered PEPS and on the cluster size.
While for $D=2$ the lowest energy is already attained with CU$_{1}$, for $D=4$ the energy improves when larger clusters are used and the FU value is obtained with CU$_{4}$.

\begin{figure}[H]
\centering
\includegraphics[width=0.63\textwidth]{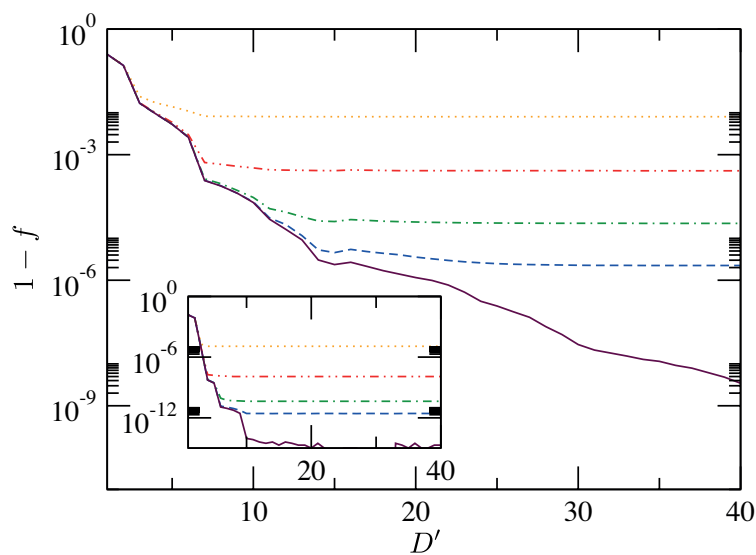}
\caption{\label{fig:Fid}
                Distance to the exact boundary MPO, $1-f$ with $f:=|\tr(\rho_{ref}^{\dag} \rho(D'))|$, for the left-most column boundary MPO $\rho(D')$ of a $D=2$ (inset) and $D=4$ (main) SU Heisenberg PEPS on a $20 \times 20$ lattice.
                We compare the cluster contraction based on clusters of size $\delta=1$ (dotted), $2$ (dash-double-dotted), $3$ (dash-dotted), and $4$ (dashed), to the full contraction (solid).
                The reference boundary MPO $\rho_{ref}$ comes from the full contraction with $D'=100$, and we adopt the normalization $\tr(\rho^{\dag} \rho)=1$.
               }
\end{figure}

This behavior agrees with our previous observation that larger bond dimensions benefit more from accurate environment representations.
We can gain further insight into this feature by looking at the convergence of a boundary MPO as function of $D'$ for different cluster sizes.
In figure \ref{fig:Fid} the environment MPO for the leftmost column, computed with different cluster sizes, is compared to the full contraction of the $L-1$ right columns with large enough $D'$, for PEPS with bond dimensions $D=2$ and $4$ on a $20 \times 20$ lattice.
We find that for each cluster size $\delta$ there exists a maximum value of $D'$ beyond which the distance to the reference boundary MPO does not decrease anymore, and that this value is smaller than the largest possible $D' = D^{2 \delta}$.
Considering a sufficiently large fixed $D'$, the distance drops exponentially with increasing cluster size until the value of the full contraction is reached.
Beyond this, larger clusters have no effect.
Finally, we can directly see that, in order to have the same distance, the $D=4$ PEPS requires larger clusters and larger boundary bond dimensions than the $D=2$ PEPS, which explains why it responds stronger to a better environment representation.

\subsection{Computation of expectation values}

\begin{figure}[H]
\centering
\includegraphics[width=0.63\textwidth]{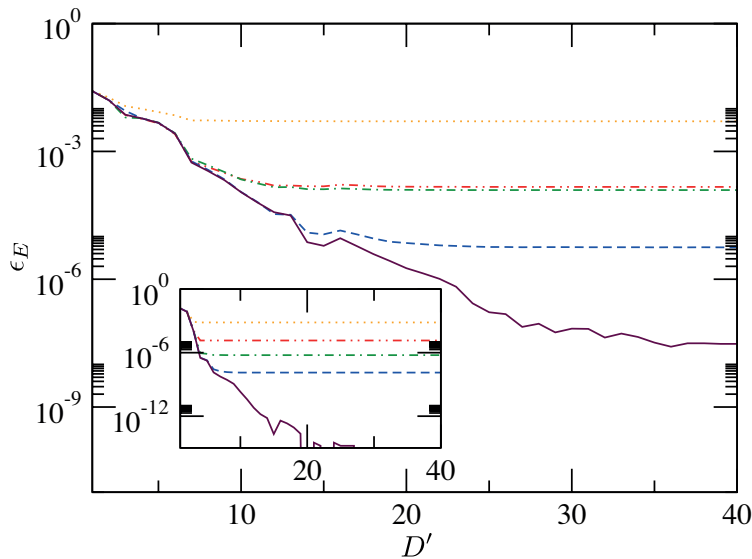}
\caption{\label{fig:CEXP}
                Relative energy error $\epsilon_{E} := |E(D')-E_{ref}|/|E_{ref}|$ of the $D=2$ (inset) and $D=4$ (main) PEPS from figure \ref{fig:Fid}, for the same setting.
                The reference value $E_{ref}$ comes from the full contraction with $D'=100$, and the clusters are formed around the individual terms of the Hamiltonian independently of each other.
               }
\end{figure}

Although we introduced clusters in the specific context of environment approximations for the tensor update, figure \ref{fig:Fid} suggests that, in fact, the reduced density matrix of any part of a PEPS can be accurately approximated by a cluster around that part, with a precision determined by the cluster size.
Therefore the cluster strategy could also be applied to the evaluation of (local) expectation values, without the need for an accurate contraction of the full TN.
To validate this idea, we computed the energy of PEPS with $D=2$ and $4$ on a $20 \times 20$ lattice using clusters of different sizes around the local terms of the Hamiltonian, shown in figure \ref{fig:CEXP}.
We observe, analogously to figure \ref{fig:Fid}, that for each cluster size the energy error converges for a certain value of $D'$, and that the larger bond dimensions require larger clusters and larger values of $D'$.
Most remarkably, we find again that for large enough fixed $D'$ the error drops exponentially with the cluster size.

\subsection{Applicability to a parallel PEPS code and to iPEPS}

In the context of finite PEPS, considered in this study, the computational cost of the environment approximation for CU$_{\delta}$ is lower than that for the full contraction only when $\delta = 0$ ($\Or(dD^{5})$) and when $\delta = 1$ ($\Or(dD^{5}D'^{2})$).
Indeed, if the boundary MPO has bond dimension $D_{0}'$, contracting it with a PEPS row and approximating the result by a new boundary with bond dimension $D_{1}'$ needs $\Or(dD^{6}D_{0}'D_{1}')+\Or(D^{4}D_{0}'^{2}D_{1}')+\Or(D^{4}D_{0}'D_{1}'^{2})$ operations. 
If $D_{0}' = D_{1}'$, we recover the scaling of the full contraction, so that the CU only results in a cheaper scheme if the environment bond dimensions $D_{i}'$ decrease as we increase the distance $i$ to the row (or column) to be updated.
Moreover, after every update of a row, the complete cluster surrounding the next row has to be contracted, without being able to reuse the previously obtained cluster boundary MPO.
This situation is different from the FU, where, when moving to the update of a new row, only one new boundary MPO has to be determined, as the previously computed and properly stored boundaries can be reused.
In the CU, the only boundary MPO that can be reused are the previously obtained separable ones, and $2\delta + 1$ new MPO have to be computed for the update of the next row.
Of those, one is separable and thus determined with computational cost $\Or(dD^{5})$, and two require $\Or(dD^{5}D'^{2})$ operations, which we can neglect, such that $2\delta - 2$ new boundary MPO have to be found with cost $\Or(dD^{6}D'^{2})+\Or(D^{4}D'^{3})$.
On the other hand, the CU takes up less memory than the FU.
The separable boundary MPO do not have to be written to hard disk but can be stored in main memory since they take up much less memory than MPO with bond dimensions $D'>1$, and then the cluster boundary MPO are computed on the fly.

Although the CU$_{\delta}$ with cluster sizes $\delta > 1$ does not reduce the computational cost of a sequential algorithm, in which one tensor is updated after another, it can reduce the cost of a parallel algorithm, in which different rows or columns are updated simultaneously on different processors.
Assuming that the time for the optimization of a boundary MPO (for a middle row) is $t_{B}$ on average, and that the update of all the tensors in a row or column is achieved in the time $t_{U}$, then the sequential FU requires $2(L-2) \cdot t_{B} + L \cdot t_{U}$ for one set of Trotter gates.
In contrast, each row or column update with the CU$_{\delta}$ for $\delta \geq 1$ necessitates the computation of only $2(\delta - 1)$ boundary MPO and thus has the cost $2(\delta-1) \cdot t_{B} + t_{U}$.
We conclude that a parallel CU algorithm can attain a $L / \delta$ speed-up.
Since $\delta$ does not depend on the system size, $L$, but only on the bond dimension, $D$, it can be chosen much smaller than $L$, such that this speed-up factor may be large.
This estimation neglects all computations with sub-leading costs $\Or(dD^{5})$ and $\Or(dD^{5}D'^{2})$ and the communication between the parallel processors.
Although the latter will have an impact on the final performance of the algorithm, we expect the speed-up to be still significant, given the fact that just the small individual tensors of separable boundary MPO have to be exchanged between different processors after each set of Trotter gates.

\begin{figure}[H]
\centering
\includegraphics[width=0.63\textwidth]{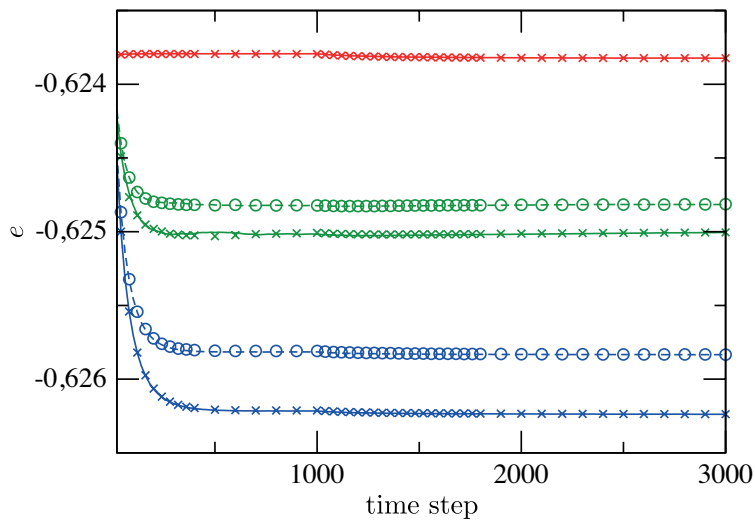}
\caption{\label{fig:PCU}
               Energy per site $e$ of the Heisenberg model on a $10 \times 10$ lattice during imaginary time evolution of a $D=4$ PEPS via the CU with parallel (lines) and sequential (symbols) tensor updates.
               The boundary MPO have fixed bond dimension $D'=1$ (top red lines and symbols), $D'=2$ (middle green lines and symbols), and $D'=16$ (bottom blue lines and symbols), and we further distinguish cluster size $1$ (dashed lines and circles) and $2$ (solid lines and crosses).
               The state propagates $1000$ time steps with $\tau = 0.01$ and then $2000$ time steps with $\tau = 0.001$.
              }
\end{figure}

The success of this parallelization strategy relies heavily on the simultaneous update of tensors in different rows.
As described in section \ref{sec:PEPS}, each tensor update is based on solving a system of linear equations that arises from the minimization of a cost function for the whole PEPS by utilizing an ALS scheme.
In this scheme one sweeps over the tensors and for each one minimizes the cost function under the assumption that all the others are fixed.
This guarantees a non-increasing cost function only when the tensors are updated sequentially.
An important question is then whether the convergence of the energy in imaginary time is as fast with the independent updates as with the sequential ones.
That this is indeed the case can be gathered from figure \ref{fig:PCU}.
The plot demonstrates an impressive agreement, which can be attributed to a minor modification of the tensor when the action of a time evolution gate is approximated in a sequential update.
We conclude that imaginary time evolution with the CU constitutes a natural basis for a parallel ground state search algorithm based on PEPS.
A similar agreement as in figure \ref{fig:PCU} cannot be expected in direct energy minimization, where a tensor is changed significantly during an update
\footnote{In direct energy minimization, the energy $\langle \psi | \hat{H} | \psi \rangle / \langle \psi | \psi \rangle$ is minimized directly by sweeping over the tensors with an ALS scheme.
This algorithm converges typically within much fewer sweeps over the PEPS than imaginary time evolution, when all sweeps for all time steps are taken into account, and therefore modifies a tensor considerably in an update.}.

Although we carried out our analysis in the framework of finite PEPS, it is clear that the CU$_{\delta}$ can also be applied to iPEPS, to replace the costly computation of the environment via the dominant boundary eigenvector with $D' > 1$ \cite{CiraciPEPS}, fixpoint corner transfer matrices \cite{OrusCTM}, or second renormalized environment \cite{XiangSRG}.
A CU procedure would only require the search for the dominant boundary eigenvector with $D'=1$, which needs $\Or(dD^{5})$ operations, followed by a cluster contraction as in the finite case.
Then, the cost and precision of both tensor update and expectation value computation would be determined by the cluster size and the bond dimension $D'$ employed in the contraction of the cluster.
Furthermore, it is always possible to evaluate clusters by means of Monte Carlo sampling \cite{VidalMC, CiracMC}.
This method requires only the contraction of the PEPS coefficients, computationally less costly than the contraction of the PEPS sandwich, for different sampled values of the physical indices $s_{1}, s_{2}, \ldots$, depicted in figure \ref{fig:PEPS}.
While a full infinite PEPS cannot be sampled, since this would necessitate determining infinitely many classical spin values, clusters open the door to variational Monte Carlo in the realm of iPEPS.
For the sampling of an observable as well as of an energy gradient, a cluster would be formed around the considered tensors and then only the physical indices of that cluster would have to be sampled.
Larger clusters necessitate longer sampling times such that the error of a finite cluster could be adjusted together with the Monte Carlo error according to the available computational resources.

\section{Conclusions}
\label{sec:Conclusions}

In this article, we have analyzed the environment representation in previous proposals, namely the Simple Update and the Single-Layer algorithm.
We have shown how the different approximations applied to the environment explain the limitations of each method in the achievable ground state accuracy, an issue that we have studied quantitatively in the context of finite PEPS.
Based on this deeper understanding, we have formulated a new proposal, the cluster strategy, that allows a systematic increase of the environment precision from the simplest possible representation, in the SU, to the most accurate full contraction, in the FU.

In its simplest form, CU$_{0}$ provides an explanation for the Simple Update in terms of a separable boundary approximation, and constitutes a slightly improved version of the latter for the models analyzed here, characterized by the same computational cost.
The first non trivial Cluster Update, CU$_{1}$, produces significantly better ground state energies than the SU, and has a lower computational cost than the FU.
In general, CU$_{\delta}$ interpolates naturally between the SU and the FU.
We have shown that increasing the cluster size improves the precision of the environment approximation exponentially.
This improvement applies directly to the computation of local observables, which can always be accelerated with the help of clusters.

Our analysis of the computational costs of the CU revealed that in the sequential update of finite PEPS any cluster size $\delta > 1$ exceeds the cost of the FU, which can reuse intermediate calculations more efficiently.
However, the CU$_{\delta}$ forms the basis of a very promising parallel PEPS algorithm, with a prospective large gain in computational time also for larger clusters.
Although our numerical studies have all been carried out in the framework of finite PEPS, we have also argued how the CU$_{\delta}$ is straightforwardly useful for the infinite iPEPS ansatz.

In summary, we have shown that the environment approximation is a key ingredient to the precision of any PEPS contraction, whether we are interested in the norm, or in some expectation value.
The CU$_{\delta}$ provides the means to control this approximation accuracy and can be used in any contraction.
It is then reasonable to think of its potential applicability to other PEPS algorithms.

\ack

ML is grateful for discussions with B Bauer, P Corboz, G De las Cuevas, S Iblisdir, V Murg, R Or\'{u}s, I Pi\v{z}orn, M Rizzi, and L Wang.
This work was partially funded by EU through SIQS grant (FP7 600645) and DFG (Cluster of Excellence NIM).
We also thank the Pedro Pascual Benasque Center for Science (CCBPP), where part of this project was carried out.

\newpage

\appendix

\section{}

Here, we list explicitly a selection of precise values, as they are used in the main text.

Regarding the reference energies, for $L=4$ the exact energy values come from exact diagonalization.
The exact Heisenberg ground state energy per site on a $4 \times 4$ lattice reads $-0.57432544$.
For $L=10$, we computed the exact values with the Quantum Monte Carlo loop algorithm from ALPS \cite{ALPS1}-\cite{ALPS3}, and we use the result for temperature $T=0.0001$, where we have checked consistency with $T=0.01$ and $0.001$.
That energy per site for a $10 \times 10$ system reads $-0.628655(2)$.

\begin{table}[H]
 \centering
 \caption{\label{tab:ESU}
                 Energy per site $e$ of the Heisenberg model on a $4 \times 4$ and $10 \times 10$ lattice, obtained by means of the SU, presented in figure \ref{fig:DESU}.
                }
 \begin{indented}
 \item[]\begin{tabular}{lll}
 \br
  $D$ & $4 \times 4$ & $10 \times 10$ \\
 \mr
  2 & -0.54404(1) & -0.61281(1) \\
  3 & -0.55396(2) & -0.61846(2) \\
  4 & -0.56281(1) & -0.62382(1) \\
  5 & -0.56628(2) & -0.62520(2) \\
  6 & -0.56684(3) & -0.62541(2) \\
  7 & -0.56696(2) & -0.62537(2) \\
  8 & -0.56715(3) & -0.62538(2) \\
 \br
 \end{tabular}
 \end{indented}
\end{table}

\begin{table}[H]
 \centering
 \caption{\label{tab:ECU0}
                 Energy per site $e$ of the Heisenberg model on a $4 \times 4$ and $10 \times 10$ lattice, obtained by means of the CU$_{0}$, presented in figure \ref{fig:DEL4L10}.
                }
 \begin{indented}
 \item[]\begin{tabular}{lll}
 \br
  $D$ & $4 \times 4$ & $10 \times 10$ \\
 \mr
  2 & -0.54404(3) & -0.61280(2) \\
  3 & -0.55397(2) & -0.61846(2) \\
  4 & -0.56287(5) & -0.62382(2) \\
  5 & -0.56637(2) & -0.62521(2) \\
  6 & -0.56694(2) & -0.62541(2) \\
  7 & -0.56706(3) &  \\
 \br
 \end{tabular}
 \end{indented}
\end{table}

\begin{table}[H]
 \centering
 \caption{\label{tab:ECU1}
                 Energy per site $e$ of the Heisenberg model on a $4 \times 4$ and $10 \times 10$ lattice, obtained by means of the CU$_{1}$ with $D'=D^{2}$, presented in figure \ref{fig:DEL4L10}.
                }
 \begin{indented}
 \item[]\begin{tabular}{lll}
 \br
  $D$ & $4 \times 4$ & $10 \times 10$ \\
 \mr
  2 & -0.54458(2) & -0.61310(2) \\
  3 & -0.5605(3) & -0.62007(1) \\
  4 & -0.56999(3) & -0.62583(2) \\
  5 & -0.57238(7) & -0.62667(2) \\
  6 & -0.57153(7) & -0.6264(2) \\
  7 & -0.57194(1) &  \\
 \br
 \end{tabular}
 \end{indented}
\end{table}

\begin{table}[H]
 \centering
 \caption{\label{tab:EFU}
                 Energy per site $e$ of the Heisenberg model on a $4 \times 4$ and $10 \times 10$ lattice, obtained by means of the FU, presented in figure \ref{fig:DEL4L10}.
                }
 \begin{indented}
 \item[]\begin{tabular}{llll}
 \br
  $D$ & $D'$ & $4 \times 4$ & $10 \times 10$ \\
 \mr
  2 & 4 & -0.54458(2) & -0.61310(2) \\
     & 8 & -0.54458(2) & -0.61310(2) \\
  3 & 9 & -0.56101(2) & -0.62002(2) \\
     & 18 & -0.5612(1) & -0.62000(2) \\
  4 & 16 & -0.5738(3) & -0.62636(3) \\
     & 32 & -0.5739(2) & -0.62637(2) \\
  5 & 25 & -0.57408(1) & -0.62732(4) \\
     & 50 & -0.57410(3) & -0.62739(1) \\
  6 & 36 & -0.57418(2) & -0.62751(2) \\
     & 72 & -0.57419(1) & -0.62770(7) \\
  7 & 49 & -0.57408(1) &  \\
     & 98 & -0.57419(1) &  \\
     & 130 & -0.57426(1) &  \\
 \br
 \end{tabular}
 \end{indented}
\end{table}

\begin{table}[H]
 \centering
 \caption{\label{tab:ECUL10D2}
                 Energy per site $e$ of the Heisenberg model on a $10 \times 10$ lattice, from $D=2$, presented in figure \ref{fig:DECUL10}.
                }
 \begin{indented}
 \item[]\begin{tabular}{lll}
 \br
  $D'$ & CU$_{1}$ & FU \\
 \mr
  1 & -0.61280(2) & -0.61280(2) \\
  2 & -0.61290(2) & -0.61289(1) \\
  3 & -0.61307(2) & -0.61307(1) \\
  4 & -0.61310(2) & -0.61310(2) \\
  100 &                   & -0.61310(2) \\
 \br
 \end{tabular}
 \end{indented}
\end{table}

\begin{table}[H]
 \centering
 \caption{\label{tab:ECUL10D4}
                 Energy per site $e$ of the Heisenberg model on a $10 \times 10$ lattice, from $D=4$, presented in figure \ref{fig:DECUL10}.
                }
 \begin{indented}
 \item[]\begin{tabular}{llllll}
 \br
  $D'$ & CU$_{1}$ & CU$_{2}$ & CU$_{3}$ & CU$_{4}$ & FU \\
 \mr
  1 & -0.62382(2) & -0.62382(2) & -0.62382(2) & -0.62382(2) & -0.62382(2) \\
  2 & -0.62481(1) & -0.62501(1) & -0.62506(5) & -0.62504(1) & -0.62508(4) \\
  4 & -0.62513(3) & -0.62583(4) & -0.62600(1) & -0.62607(2) & -0.62602(1) \\
  12 & -0.62583(2) & -0.62623(2) & -0.62631(2) & -0.62634(3) & -0.62635(2) \\
  16 & -0.62583(2) & -0.62623(2) & -0.62632(3) & -0.62635(3) & -0.62636(3) \\
  20 &                       & -0.62624(2) & -0.62632(2) & -0.62635(2) & -0.62636(2) \\
  32 &                       &                       &                       &                        & -0.62637(3) \\
 \br
 \end{tabular}
 \end{indented}
\end{table}

\newpage

\section{}

We used the following setup for time evolution and energy computation, if not explicitly stated otherwise.

We initialize imaginary time evolution with a separable $D=2$ PEPS in which the zeroes are replaced by noise as uniformly distributed random numbers from $[-0.01, 0.01]$.
This state is evolved for $N_{1}$ steps with $\tau_{1}$, followed by $N_{2}$ steps with $\tau_{2}$, and so on, what we abbreviate to the short notation $(N_{1} \times \tau_{1}, N_{2} \times \tau_{2}, \ldots )$ for fixed bond dimension $D$.
In order to specify a successively growing value of $D$, we introduce the recursive notation $(D_{i+1}=D_{i}^{\tau}+1, N_{1} \times \tau_{1}, N_{2} \times \tau_{2}, \ldots )$.
It defines the next PEPS for the propagation with bond dimension $D_{i+1}$ as the final state of the previous evolution with bond dimension $D_{i}$ and time step $\tau_{i}$ with a by $1$ incremented bond dimension.
In the case of the Cluster and Full Update, the additional parameter $D'$ is typically chosen as $D'=1$, $2$, and, related to $D$, as $D'=D$, $D^{2}$, and so on.
The final PEPS obtained with a certain value of $D'$ is always the initial state for increased $D$ with that $D'$.

Regarding the energy computation, all energies are evaluated with $D'=100$ for the final PEPS corresponding to the smallest time step.
We define the energy error as the difference between the energy of this final state and the energy of an intermediate state.
The latter is either the PEPS obtained after half of the evolution or the final PEPS corresponding to the immediately larger time step, depending on wether or not the propagation was also performed with this larger time step.
\\

\noindent\textbf{Figure \ref{fig:DESU}}:\\
We propagate the initial $D=2$ PEPS $1000$ time steps with $\tau=0.1$, then $2000$ time steps with $\tau=0.01$, then $8000$ time steps with $\tau=0.001$, and then according to the configuration $(D_{i+1}=D_{i}^{\tau=0.01}+1, 2000 \times \tau=0.01, 8000 \times \tau=0.001)$.
\\

\noindent\textbf{Figure \ref{fig:DNSL}}:\\
We propagate the initial $D=2$ PEPS each time $10000$ steps first with $\tau=0.1$, then with $\tau=0.01$, then with $\tau=0.001$, then $20000$ steps with $\tau=0.0001$, and finally $50000$ steps with $\tau=0.00001$.
The $D=3$ and $4$ results, used in the analysis, were obtained by evolving the final $D=2$ PEPS from $\tau=0.0001$ further according to the configuration $(D_{i+1}=D_{i}^{\tau=0.0001}+1, 20000 \times \tau=0.0001)$.
\\

\noindent\textbf{Table \ref{tab:GSU}}:\\
We apply the SU to a $11 \times 11$ Ising model with different magnetic fields $B$, and propagate the initial $D=2$ PEPS $10000$ time evolution steps with $\tau=0.1$ and then $10000$ steps with $\tau=0.01$.
The resulting PEPS is further evolved with the CU$_{0}$ for $10000$ time steps with $\tau = 0.01$.
All shown numbers are converged to machine precision.

\newpage

\noindent\textbf{Figure \ref{fig:DEL4L10}} (a) ($N = 4 \times 4$):\\
The initial state of the imaginary time evolution is the converged $D=2$ SU ground state approximation to time step $\tau = 0.01$.
We propagate this state $1000$ time steps with $\tau=0.01$, then $2000$ time steps with $\tau=0.001$, and then according to the configuration $(D_{i+1}=D_{i}^{\tau=0.01}+1, 1000 \times \tau=0.01, 2000 \times \tau=0.001)$ up to bond dimension $D=5$.
Then, we continue with $(D_{i+1}=D_{i}^{\tau=0.01}+1, 500 \times \tau=0.01, 1000 \times \tau=0.001)$.
In the case of the FU with $D=7$ and $D'=130$, the state propagates $100$ time steps with $\tau = 0.01$.
\\

\noindent\textbf{Figure \ref{fig:DEL4L10}} (b) ($N = 10 \times 10$):\\
The initial state of the imaginary time evolution is the converged $D=2$ SU ground state approximation to time step $\tau = 0.01$.
We propagate this state $1000$ time steps with $\tau=0.01$, then $2000$ time steps with $\tau=0.001$, and then according to the configuration $(D_{i+1}=D_{i}^{\tau=0.01}+1, 1000 \times \tau=0.01, 2000 \times \tau=0.001)$.
In the cases of the CU$_{0}$ and the CU$_{1}$ we use this time evolution configuration up to $D=5$, and for $D=6$ propagate the states $500$ time steps with $\tau = 0.01$ and then $500$ time steps with $\tau = 0.001$.
In the case of the FU we use this configuration up to $D=4$, and for $D=5$ evolve the states $500$ time steps with $\tau = 0.01$ and then $1000$ time steps with $\tau = 0.001$, and for $D=6$ we propagate the states $500$ time steps with $\tau = 0.01$.
\\

\noindent\textbf{Figure \ref{fig:DECUL10}}:\\
The initial state of the imaginary time evolution is the converged SU ground state approximation to time step $\tau = 0.01$ with the considered bond dimension.
In the case of $D=2$, we propagate this state $1000$ time steps with $\tau=0.01$, then $2000$ time steps with $\tau=0.001$.
In the case of $D=4$, for CU$_{1}$ we use $(1000 \times \tau=0.01, 2000 \times \tau=0.001)$, for CU$_{2}$ $(1000 \times \tau=0.01, 1000 \times \tau=0.001)$, and for CU$_{3}$, CU$_{4}$, and FU we use $(500 \times \tau=0.01, 1000 \times \tau=0.001)$.

\end{document}